\documentstyle[11pt,moriond,epsfig]{article}

\def\laq{\raise 0.4ex\hbox{$<$}\kern -0.8em\lower 0.62
ex\hbox{$\sim$}}
\def\gaq{\raise 0.4ex\hbox{$>$}\kern -0.7em\lower 0.62
ex\hbox{$\sim$}}

\def\NPB{{\em Nucl. Phys.} B}
\def\PLB{{\em Phys. Lett.}  B}
\def\PRL{\em Phys. Rev. Lett.}
\def\PRD{{\em Phys. Rev.} D}

\def\be{\begin{equation}}
\def\ee{\end{equation}}
\def\bea{\begin{eqnarray}}
\def\eea{\end{eqnarray}}

\begin{document}
\vspace*{4cm}
\title{COSMIC GRAVITATIONAL WAVE BACKGROUND \\ IN STRING COSMOLOGY}

\author{ RAM BRUSTEIN }

\address{Department of Physics, Ben-Gurion University, 
Beer-Sheva 84105, Israel}

\maketitle\abstracts{
String cosmology models predict a stochastic cosmic background of 
gravitational waves with a characteristic spectrum. I describe the background,
present astrophysical and cosmological bounds on it, and outline
how it may be possible to detect it with gravitational wave detectors. }
\section{Introduction}
A robust prediction of models of string cosmology which realize the
pre-big-bang scenario \cite{sfd,pbb} is that our
present-day universe contains a cosmic gravitational wave background 
\cite{bggv,gg,gwg},  with a spectrum which is quite different than that predicted
by many other early-universe  cosmological
models \cite{grishchuk,turner,myreview,maggiore}.  In the pre-big-bang scenario the
evolution of the universe starts from a state of very small curvature and
coupling and then undergoes a long phase of dilaton-driven kinetic inflation
reaching roughly Planckian energy densities \cite{exit1},  and at some later time joins
smoothly standard  radiation dominated cosmological evolution, thus giving rise
to a singularity free inflationary cosmology. 

It is the purpose of this talk to explain the properties of the cosmic 
gravitational  wave background predicted by string cosmology, to explain astrophysical 
and cosmological bounds  on its shape and strength and to show that currently
operating and planned gravitational  wave detectors could further constrain the
spectrum and perhaps even detect it. The emphasis  is on the general properties
of the spectrum and   the basic
experimental setup needed for its detection, rather than on precise numerical 
estimates and technical details.

Because the gravitational interaction is so weak, a 
background of gravitational radiation decouples from the matter in the
universe  at very early times and carries with it a picture of the state of the
universe  when energy densities and temperatures were extreme.
The weakness of the gravitational interaction makes a detection of such a
background very hard, and necessitates a strong signal. String
cosmology provides perhaps the strongest source possible: the whole universe, 
accelerated to roughly Planckian energy densities. A discovery of any primordial
gravitational  wave background, and in particular, the one predicted by string
cosmology,  could therefore provide unrivaled exciting information on the
very early universe. 
\section{Cosmic gravitational wave background in string cosmology}
\label{section:first}
In models of string cosmology \cite{bggv}, the universe passes through
two early inflationary stages.  The first of these is called the
``dilaton-driven" period and the second is the ``string" phase.  Each
of these stages produces stochastic gravitational radiation by  the standard
mechanism of amplification of quantum fluctuations \cite{mukh}. 
Deviations from homogeneity and isotropy of the metric field are generated by 
quantum  fluctuations around the homogeneous and isotropic background, and then
amplified by the accelerated expansion of the universe. The transverse and 
traceless part of these fluctuations are the gravitons. In practice, we
compute graviton production by solving  linearized
perturbation equations with vacuum fluctuations boundary conditions. 
The production strength of gravitons depends on the curvature and coupling. Since
at the end of the accelerated expansion phase curvatures reach the 
string curvature, and the coupling reaches approximately today's coupling, graviton
production is expected to be at the strongest possible level.

In order to describe the background of gravitational radiation, it is
conventional to use a spectral function 
$
\Omega_{\rm GW}(f) = {1 \over \rho_{\rm critical}} {d \rho_{\rm
GW} \over d \ln f}.
$ Here $ d \rho_{\rm GW}$ is the (present-day) energy density in
stochastic gravitational waves in the frequency range $ d \ln f$, and
$\rho_{\rm critical}$ is the critical energy-density required to just
close the universe,
$
\rho_{\rm critical} = { 3 c^2 H_0^2 \over 8 \pi G} \approx 1.6 \times
10^{-8} {\rm h}_{100}^2 \rm \> ergs/cm^3,
$
where the Hubble expansion rate $H_0$ is the rate at which our universe
is currently expanding,
$
H_0 = {\rm h}_{100} \> 100 \> {\rm  Km \over sec-Mpc} = 
3.2 \times 10^{-18}{\rm h}_{100} {\rm 1 \over sec}.
$
 ${\rm h}_{100}$  is believed to lie in the
range $0.5 < {\rm h}_{100} < 0.8$. The spectral function is related to the
dimensionless strain $h$, 
$\Omega_{\rm GW}(f) \simeq 10^{36} h_{100}^{-2}(f/ {\rm Hz})^2 h(f)^2$ 
and to the strain in units $1/\sqrt{Hz}$, $\sqrt{S_h(f)}$, 
$\Omega_{\rm GW}(f) \simeq 10^{36} h_{100}^{-2} (f/ {\rm Hz})^3 S_h(f)$.

The spectrum of gravitational radiation produced in the dilaton-driven
(and string) phase was estimated in [3],
$
\Omega_{GW}(f) \simeq
z_{eq}^{-1} g_s^2
 \left(f \over f_S\right)^3
\left[ 1+ z_S^{-3} \left(g_1\over g_S\right)^{2}\right],  f<f_S,
$
where some logarithmic correction factors were dropped.
The coupling $g_1$ is today's
coupling, assumed to be constant from the end of the string phase until today,
$g_S$ is the coupling at the end of the dilaton-driven phase, and $f_S$ is the
frequency marking the end of the dilaton-driven phase. The frequency $f_1=f_S
z_S$ is the frequency at the end of the string phase, where $z_S$ is the total
redshift during the string phase and $z_{eq}\sim 10^4$ is the redshift from 
matter-radiation equality until today. The spectrum can be expressed in a more
symmetric  form \cite{sduality},    
$
\Omega_{GW}(f)
\simeq z_{eq}^{-1} g_1^2 \left(\frac{f}{f_1}\right)^3  
 \left[ z_S^3 (g_S/g_1)^2+ z_S^{-3} (g_S/g_1)^{-2}\right].
$
  Note that the spectrum is
invariant under the exchange  $z_s^3 (g_s/g_1)^2 \leftrightarrow z_s^{-3}
(g_s/g_1)^{-2}$
 and that this implies a lower bound on the  spectrum,  
$
\Omega_{GW}(f) \gaq
2 z_{eq}^{-1} g_1^2 \left(\frac{f}{f_1}\right)^3.$
The lower bound is obtained for the ``minimal spectrum" 
with $z_S=1$ and $g_S/g_1=1$ describing a cosmology with almost no intermediate
string phase.
\begin{figure}
%\begin{center}
\psfig{figure=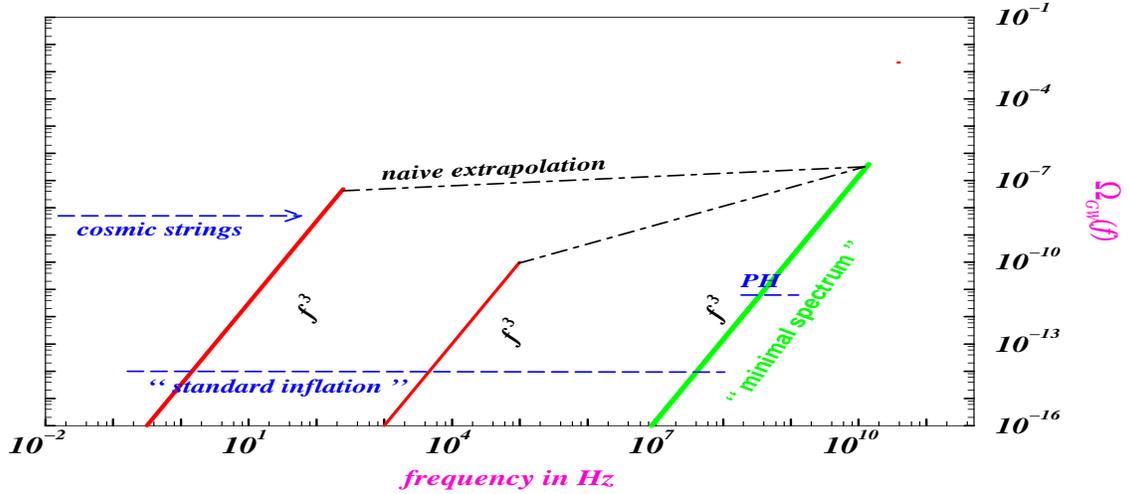,width=4.5in,height=4.2in}
%\end{center}
\vspace{-1.6in}
\caption{Spectrum of GW background. The minimal spectrum discussed in the text and two
other possible spectra are shown. Also shown are estimated spectra in other cosmological
models.\hfil  
\label{fig:ogw} }
\end{figure}

In the simplest model, which we will use to estimate the spectrum and prospects 
for its detection, the spectrum depends upon four parameters.   The first pair
of parameters  are the maximal frequency $f_1$  above which gravitational
radiation is not produced and $g_1$, the coupling at the end of the string
phase. The second pair of these are $z_S$ and $g_S$. The second pair of
parameters can be traded for  the  frequency ${f_{\rm S}}=f_1/z_S$ and the
fractional energy density $\Omega^{\rm S}_{\rm GW}=\Omega_{\rm GW}(f_S)$  
produced at the end of
the dilaton-driven phase. At the moment, we cannot compute $g_S$ and $z_S$ from
first principles, because they involve knowledge of the evolution during the
high curvature string phase. We do, however, expect $z_S$  to be quite large.
Recall that $z_S$ is the total redshift during the string phase, 
 and that during this phase 
the curvature and expansion rate are  approximately string scale,
therefore, $z_S$ grows roughly exponentially with the duration (in string times)
of this phase. Some particular exit models \cite{exit2} suggest
that $z_S$ could indeed be quite large. I
cannot estimate, at the moment, a likely range for the ratio $g_1/g_S$  except
for the reasonable assumption $g_1/g_S>1$.

A useful approximate form for the spectrum in the range $z_S>1$ and
$g_1/g_S\gaq 1$  is the following~\cite{rb}
\be \label{e:approx}
\Omega_{\rm GW}(f)=
\cases{
{\Omega^{\rm S}_{\rm GW} (f/{f_{\rm S}})^3}  &  { $f<{f_{\rm S}}$} \cr
 & \cr
{\Omega^{\rm S}_{\rm GW} (f/{f_{\rm S}})^{\beta} } &   ${f_{\rm
S}}<f<f_1$ \cr
 & \cr
{0} &   $f_1<f$}
\ee
where
$
\beta=\frac{\log\left[\Omega_{\rm GW}(f_1)/
\Omega^{\rm S}_{\rm GW}\right]}{\log\left[f_1/{f_{\rm S}}\right]}
$
is the logarithmic slope of the spectrum produced in the string phase 
(see also other models \cite{spgravitons}).
If we assume that there is no late entropy production and make
reasonable choices for the number of effective degrees of freedom, then
two of the four parameters may be determined in terms of the Hubble
parameter $H_{\rm r}$ at the onset of radiation domination immediately
following the string phase of expansion \cite{peak},
$
f_1= 1.3 \times 10^{10} \> {\rm Hz} \left( { H_{\rm r} \over 5 
\times 10^{17} \> {\rm GeV}} \right)^{1/2}
$
and
$
\Omega_{\rm GW}(f_1) = 1 \times 10^{-7} {\rm h}_{100}^{-2} 
\left( { H_{\rm r} \over 5 \times 10^{17} \> {\rm GeV}} \right)^2.
$
Spectra for some arbitrarily chosen parameters and possible backgrounds
from other cosmological models are shown in Fig.~\ref{fig:ogw}.  
The label PH denotes  preheating after inflation \cite{tkach}.
\section{Astrophysical and cosmological bounds}\label{section:second}
\begin{figure}
%\begin{center}
\psfig{figure=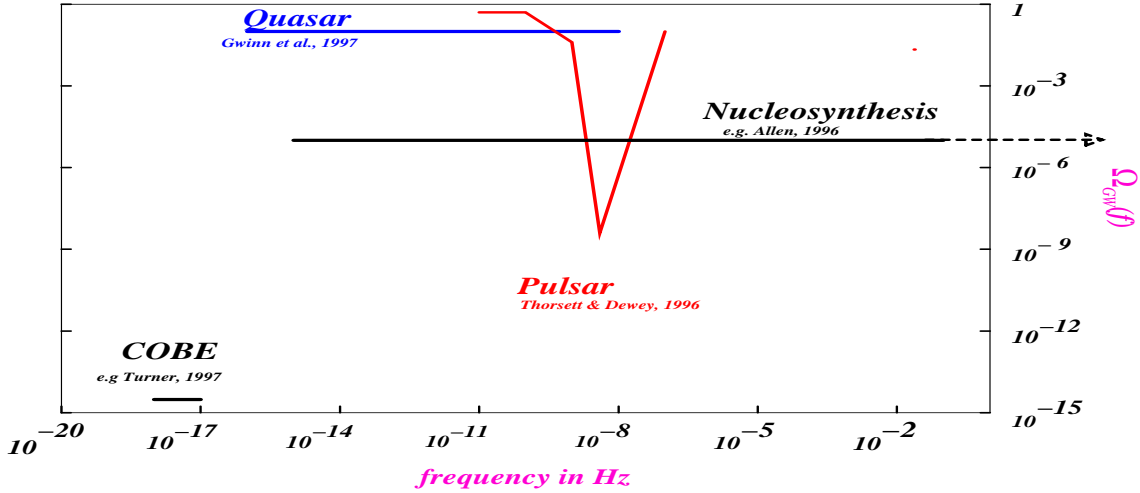,width=4.5in,height=4.2in}
%\end{center}
\vspace{-1.6in}
\caption{Cosmological and astrophysical upper bounds on the cosmic
gravitational wave background.
\label{fig:cabounds}}
\end{figure}

At the moment, the most restrictive observational constraint on the
spectral parameters comes  from  the standard model of big-bang
nucleosynthesis (NS) \cite{ns1}.  This restricts the total energy
density in gravitons to less than that of approximately one massless
degree of freedom in thermal equilibrium. This bound implies that \cite{ba}
\be
\int \Omega_{\rm GW}(f) d \ln f  = \Omega_{\rm GW}^{\rm S} \left[
\frac{1}{3}+ \frac{1}{\beta}\left( \left(f_1/f_{\rm S}
\right)^\beta-1\right)\right] < 0.7 \times 10^{-5} {\rm h}^{-2}_{100}.
\label{nucleo}
\ee
where we have assumed an allowed $N_\nu=4$ at NS, and have substituted
in the spectrum (\ref{e:approx}). The NS bound and additional cosmological 
and astrophysical bounds are shown in Fig.~\ref{fig:cabounds}, where $h_{100}$
was set to unity.

The line marked ``Quasar" in Fig.~\ref{fig:cabounds} corresponds to a bound 
coming from quasar proper motions. A stochastic background of gravity waves (GW)
makes the signal from distant quasars scatter randomly on its way to earth.
This may cause quasar proper motions. An upper bound on quasar proper
motions can be translated into an upper bound on a stochastic
background~\cite{gwinn}. A typical strain $h$ may induce  proper motion $\mu$,
$h/f\sim \mu$. The sensitivity reached was approximately microarcsecond per
year \cite{gwinn},  corresponding to a dimensionless strain of 
about $h\sim 5 \times 10^{-9}$
at frequencies below the observation time: approximately (20 years)$^{-1}\sim
5\times 10^{-9}Hz$, leading to $\Omega_{GW}\laq 0.1 h_{100}^{-2}$.

The line marked ``COBE" in Fig.~\ref{fig:cabounds} corresponds to the 
bound coming from energy density
fluctuations in the cosmic microwave background,  which can be
expressed in terms of the measured temperature fluctuations $\Delta T/T$,
$\Omega({\rm perturbations})\simeq  (\frac{\Delta T}{T})^2  \Omega_\gamma\sim
10^{-10}\times  10^{-4}=10^{-14} h_{100}^{-2}$. Since it is known \cite{turner}
that $\Omega_{GW}\laq 0.1  \Omega({\rm perturbations})$,  it follows that
$\Omega_{GW} h_{100}^2 \laq 10^{-15}$ at frequencies  $10^{-18} h_{100} Hz -
10^{-16}h_{100} Hz$.

The curve marked ``Pulsar" represents the bound coming from millisecond pulsar 
timing~\cite{td}.  Assuming known distance and signal emission times, the pulsar functions
as a giant one-arm interferometer. The statistics of pulse arrival time residuals
$\Delta T$, puts an upper bound on any kind of noise in the
system, including a  stochastic background of GW. The typical strain
sensitivity is  $h\sim \frac {\Delta T}{T}$, where $T$ is the total observation
time, reaching by now 20 years $\sim 6 \times 10^{8} sec$ and $\Delta T\sim 10
\mu s$  is the accuracy in measuring time residuals. Translated into
$\Omega_{GW}(f)$, this yields the bound shown in the figure, which is most restrictive
at frequencies $f\sim 1/T \sim 5\times 10^{-9} Hz$.
 
Notice that all the existing bounds are in the very low frequency range, while
the expected signal from string cosmology is in a higher frequency range. The
bounds are therefore not very restrictive.

 \begin{figure}
%\begin{center}
\psfig{figure=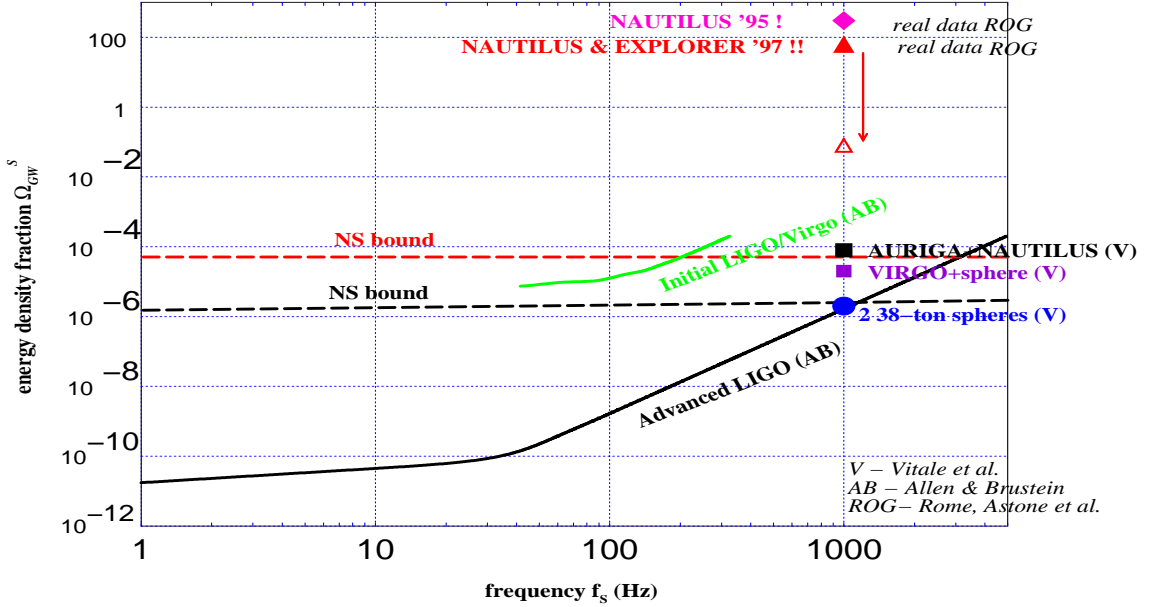,width=4.5in,height=5.0in}
%\end{center}
\vspace{-1.8in}
\caption{Detection sensitivity of relic GW by operating and planned GW 
detectors. The interesting region of parameter space is below
the ``NS bound" lines.\hfil\hfil\hfil
\label{fig:gensens}}
\end{figure}

\section{Detecting a string cosmology stochastic gravitational wave
background}\label{section:third}
A number of authors \cite{stoch,vitale,allen} have shown how one
can use a network of two or more gravitational wave antennae to detect
a stochastic background of gravitational radiation.  The basic idea is
to correlate the signals from separated detectors, and to search for a
correlated strain produced by the gravitational wave background, which
is buried in the instrumental noise.  It has been shown by
these authors that after correlating signals for time $T$ the ratio of ``Signal"
to ``Noise" (squared) is given by an integral over frequency $f$:
\be \label{e:sovern}
\left( {S \over N} \right)^2 =
{9 H_0^4 \over 50 \pi^4} T \int_0^\infty df \>
{\gamma^2 (f) \Omega_{\rm GW}^2(f) \over f^6 P_1(f) P_2(f)}.
\ee  
The instrument noise in the detectors is described by the one-sided
noise power spectral densities $P_i(f)$.  The LIGO project is building
two identical detectors,  which we will refer to as the ``initial"
detectors. These detectors will be upgraded to so-called ``advanced" detectors. 
 Since the two detectors are identical in design, $P_1(f)=P_2(f)$.  
The design goals for the detectors specify these functions \cite{science92}. 
The design  noise power spectrum for the Virgo detector \cite{virgo} and 
noise power spectral densities of operating 
and planned resonant mass GW detectors (``bars") \cite{naut,auriga} are also
known.
The overlap reduction function $\gamma(f)$ is determined by the
relative locations and orientations of the two detectors. It is
identical for both the initial and advanced LIGO detectors, and has been
determined for many pairs of GW detectors \cite{vitale,allen}.

Making use of the prediction from string cosmology (\ref{e:approx}), 
we may use equation (\ref{e:sovern}) to assess the detectability of this stochastic
background.  For any given set of parameters we may numerically
evaluate the signal to noise ratio $S/N$; if this value is greater than
$1.65$ then with at least 90\% confidence, the background can be
detected by a given pair of detectors.  The regions of detectability in parameter
space are shown in Fig.~\ref{fig:gensens}. The region below the  NS bound lines
and above the advanced LIGO curve is the region of interest. Two NS bounds are shown, 
the upper, more relaxed bound, assumes no GW production during the 
string phase \cite{ba}. The points at 1 KHz come from operating
and planned resonant mass detectors.
Some are taken from real experiments, an upper bound from a single detector run
 \cite{nautstoch}, and the first modern 12.5 hours correlation experiment
 between Nautilus and Explorer \cite{nautexplo97}. The arrow points
to a hollow triangle showing by how much the correlation 
experiment can be improved if Nautilus works properly and the experiment could 
be done for one year. Other points are from theoretical
calculations~\cite{vitale}. 
 For Fig.~3 we have assumed ${\rm h}_{100}=0.65$
and $H_{\rm r}=5 \times 10^{17} \> {\rm GeV}$. 
\section*{Acknowledgments}
I would like to thank my collaborators Bruce Allen, Maurizio Gasperini and 
Gabriele Veneziano.
Thanks to many GW experimentalists for their help and special thanks to the 
ROG collaboration and Pia Astone for access to their data. 
This work is supported in part by the  Israel
Science Foundation administered by the Israel Academy of Sciences and
Humanities.

\end{document}